\documentstyle[12pt]{article}
\begin{document}

\newfont{\gothic}{eufm10 at 12pt}
\newcommand{\g}{{\gothic g}}
\newcommand{\nn}{\nonumber}
\newcommand{\be}{\begin{equation}}
\newcommand{\ee}{\end{equation}}
\newcommand{\bea}{\begin{eqnarray}}
\newcommand{\eea}{\end{eqnarray}}
\newcommand{\Ug}{{\cal{U}}\g}
\newcommand{\Cg}{{\cal{C}}\g}
\newcommand{\Yg}{{\cal{Y}}\g}
\newcommand{\vsp}{\vspace{0.2in}}
\newcommand{\va}{\mbox{\boldmath $1$}}
\newcommand{\st}[1]{\mbox{\boldmath $#1$}}
\newcommand{\om}{\mbox{\boldmath $\omega$}}
\newcommand{\Wg}{W^{\infty /2}\g}
\newcommand{\Wl}{W\ell}
\newcommand{\Wtl}{W^{\infty /2}\tilde{\ell}}
\newcommand{\tl}{\tilde{\ell}}
\newcommand{\Sl}{S\ell'}
\newcommand{\El}{\wedge\ell'}
\newcommand{\Stl}{S^{\infty /2}\tilde{\ell}'}
\newcommand{\Etl}{\wedge^{\infty /2}\tilde{\ell}'}
\newcommand{\Be}{\beta}
\newcommand{\Ga}{\gamma}
\newcommand{\lra}{\leftrightarrow}
\newcommand{\Lra}{\Leftrightarrow}
\newcommand{\ra}{\rightarrow}
\newcommand{\mt}{\mapsto}
\newcommand{\Ki}{{\cal{K}}}
\newcommand{\la}{\leftarrow}
\newcommand{\da}{\downarrow}
\newcommand{\ua}{\uparrow}
\newcommand{\vv}{\vspace{0.1in}}
\newtheorem{thm}{Theorem}
\newtheorem{prop}{Proposition}
\newtheorem{lem}{Lemma}
\newtheorem{cor}{Corollary}

\begin{center}
{\bf SOME COHOMOLOGY OPERATORS} \\ {\bf IN 2-D FIELD THEORY}
\end{center}
\vsp

\begin{center}
F\"{U}SUN AKMAN
\\ {\em Mathematical Sciences Research Institute}
\\ {\em 1000 Centennial Drive, Berkeley, CA 94720}
\\ {\em e-mail: akman@msri.org}
\end{center}
\vsp

\begin{center}
ABSTRACT
\end{center}
\vsp

\begin{center}
{\small\parbox{5in}{It is typical for a semi-infinite cohomology
complex associated with a graded Lie algebra to occur as a vertex
operator (or chiral) superalgebra where all the standard operators of
cohomology theory, in particular the differential, are modes of vertex
operators (fields). Although vertex operator superalgebras -with the
inherent Virasoro action- are regarded as part of Conformal Field
Theory (CFT), a VOSA may exhibit a square-zero operator (often, but
not always, the semi-infinite cohomology differential) for which the
Virasoro algebra acts trivially in the cohomology. Capable of shedding
its CFT features, such a VOSA is called a ``topological chiral
algebra'' (TCA). We investigate the semi-infinite cohomology of the
vertex operator Weil algebra and indicate a number of differentials
which give rise to TCA structures.}}
\end{center}
\vsp
\vsp

{\bf 1. Introduction}
\vsp

We will always work with an algebraically closed ground field $K$ of
characteristic zero.
\vsp

{\em 1.1. Semi-infinite Cohomology}
\vsp

Designed for {\em tame} Lie algebras, that is, {\bf Z}-graded Lie
algebras
\be \g=\oplus_{n}\g_{n} \label{g} \ee
with $\dim \g_n$ finite, and for a large category of $\g$ modules,
semi-infinite cohomology made its debut in mathematics with
B. Feigin's 1984 paper [Fe]. It was studied independently by
physicists as ``BRST cohomology'', and further investigated by I.~B.
Frenkel, H. Garland, and G. J. Zuckerman in [FGZ] (see the references
therein for physics literature).
A characterization of the semi-infinite
cohomology differential as an associative algebra derivation which is
generically square-zero first appeared in [A1]. It is easier to
outline this last approach for the special (classical) case
$\g=\g_0$.
\vsp

When the Lie algebra is finite dimensional, a differential complex is
just the tensor product of a $\g$ module $M$ with the exterior algebra
on the dual of $\g$, i.e.
\be M\otimes\wedge\g' \; .   \label{M}  \ee
The universal enveloping algebra $\Ug$ acts on the first factor and
the Clifford algebra $\Cg$ on the second, hence any complex (\ref{M})
is a natural module over the associative algebra
\be \Yg = \Ug \otimes \Cg \; . \label{Y} \ee
Furthermore the adjoint action of $\g$ on $\Ug$ and the action on
$\Cg$ induced from the adjoint and coadjoint representations give rise
to a total action $\theta$ of $\g$ on $\Yg$ via inner derivations
(represented by elements $\theta(x)$ of $\Yg$ for every $x$ in $\g$).
Note also that $\Cg$ has a natural supergrading where generators from
$\g$ have degree~$-1$ and generators from $\g'$ have degree~1.
Then it can be shown that ([A1])
\vsp

\begin{thm}
There exists a unique inner derivation of $\Yg$ (represented by an
element $d$) with superdegree 1 which satisfies the {\em Cartan
identity}
\be d\iota(x)+\iota(x)d=\theta(x) \; \; \, \, \forall x \in \g
\label{Cartan} \ee
in $\Yg$.
\end{thm}
\vsp

Here $\iota(x)$ can be thought of as a generator of $\Cg$
coming from $\g$, or as the substitution (contraction) operator on
$\wedge\g'$. The derivation $d$ is square-zero and acts on any
complex~(\ref{M}), enabling us to define the (classical) cohomology of
$\g$ with coefficients in $M$. Passing to infinite dimensional $\g$
and semi-infinite cohomology requires a little more work; the only
significant differences are that one dualizes graded tame vector
spaces piecewise (all duals are thus ``restricted'') , and allows a
``completion'' of $\Yg$. The process also involves replacing the
exterior algebra with the semi-infinite exterior module (to be defined
later). A generalization of Theorem~1 holds with the
extra condition that the derivation preserves the {\bf Z}-grading
induced by that of $\g$ in the completed algebra ([A1]).
\vsp

We will assume the existence and properties of semi-infinite
structures and present only the formulas for various operators when
the appropriate notation is introduced.
\pagebreak

{\em 1.2. Vertex Operator Superalgebras}
\vsp

A {\em vertex operator superalgebra} ({\em VOSA}), or a {\em chiral
superalgebra}, is a {\bf Z}-bigraded vector space
\be V=\oplus_{j,n}V^j[n] \label{V} \ee
with additional structure. We will refer to $n$ as the {\em degree},
or {\em weight} (as physicists do). The other grading, $j$, will be
the {\em superdegree} ({\em fermion number, ghost number}...).
Algebraic properties of VOSA's have been isolated, refined, and
generalized by a number of mathematicians. The Monster book [FLM] and
the monographs [FHL], [Li], [FZ], [DL] are good references.
First of all a VOSA is equipped with an injective linear map
\bea && V \rightarrow End(V)[[z,z^{-1}]] \nn \\ &&
\st{v} \mapsto v(z)=\sum v_n z^{-n-1} \label{VO} \eea
where $\st{v}$ is called a {\em state}, $v(z)$ is called a {\em vertex
operator} ({\em VO}), or a {\em field}, and $v_n$ is a {\em mode}.
We also require that $V$ be a graded representation of the
Virasoro algebra (with respect to $n$) such that $L_0$ acts
semisimply on $V$ and has eigenvalue $n$ on $V^{\ast}[n]$.
Recall that $Vir$ has a basis consisting of
$\{ L_n \}_{n\in Z}$ and a central element $c$ with relations
\be {[L_n,L_m]}=(n-m)\, L_{n+m}+\frac{m^3-m}{12}\, \delta_{n+m,0}\, c \; .
\label{Vir} \ee
An important axiom is
\be L_{-1}\st{v} \; \lra \; \frac{d}{dz} v(z) \; .
\label{dz} \ee
Next, there are special elements
\bea && \va \; \mbox{(``vacuum'')} \; \lra \; \mbox{id}\cdot
z^0 \nn \\ && \om \; \lra \; \sum L_n z^{-n-2} \;
\mbox{(``stress-energy field'')} \label{special} \eea
with weights $0$, $2$ and superdegrees $0$, $0$ respectively.
The main axiom of a VOSA, the {\em Cauchy-Jacobi
identity} which can be found in [FLM], may be replaced by the
following ``commutativity'' condition as was shown by C. Dong and J.
Lepowsky ([DL]): For any homogeneous
$\st{v}$, $\st{w}$ in $V$ there exists $t \gg 0$ such that
\be {[v(z_1),w(z_2)]}(z_1-z_2)^t=0 \label{comm} \ee
where $[\; , \; ]$ denotes the supercommutator of two formal series.
\vsp

Any two elements $\st{v}$ and $\st{w}$ of $V$ produce infinitely
many elements via the multiplications
\be v_n \cdot \st{w} \; . \label{mult} \ee
In this sense, one may ask whether there is a set of naturally chosen
elements of $V$ that generate the whole VOSA. One usually constructs
a VOSA from finitely many generators, but it is much more difficult to
determine whether a VOSA constructed by other means is finitely
generated. Such questions may arise, for example, when the VOSA is a
cohomology.
\vsp

The coefficient of $z^{-1}$ in (\ref{VO}), namely $v_0 \in End(V)$, is
called the {\em residue}, or {\em charge}, of the VO $v(z)$.
It is well-known that
\be v_0\va =0 \label{reskillsvac} \ee
and
\be {[v_0,w(z)]}=(v_0\cdot\st{w})(z) \label{resbrac} \ee
for all $\st{v}$, $\st{w}$ in a VOSA. In particular,
\be v_0\cdot\st{w}=0 \; \mbox{in} \; V \; \Lra \;
[v_0,w_n]=0 \; \mbox{in} \; End(V). \label{rescond} \ee
Residues give rise to new VOSA's, because
\vsp

\begin{lem}
If $\om$ is in $Ker \, v_0$ for some $v\in V$, then $Ker\, v_0$ is a
vertex operator subalgebra of $V$.
\end{lem}

\begin{lem}
For a square-zero residue $v_0$, $Im\, v_0$ is an ideal of the VOSA
$Ker\, v_0$. Consequently the cohomology $H(V,v_0)$ is a VOSA
(provided that $\om$ is in $Ker\, v_0$).
\end{lem}
\vsp

Again these two are widely used results for which short proofs can be
found in [A2]. If $v_{0}^{2}=0$, we will call the state $\st{v}$ a
{\em reduction element}, inspired by Hamiltonian reduction. It is a
fact of life that most semi-infinite complexes not only show up as
VOSA's, but also manifest a special reduction element with residue
equal to the semi-infinite cohomology differential (which will be
denoted by $Q$ as is customary from now on). Examples are in [LZ].
\vsp

{\bf 2. The Vertex Operator Weil Algebra}
\vsp

{\em 2.1. Definition and Properties}
\vsp

The {\em semi-infinite Weil complex} $\Wg$ associated to a tame Lie
algebra $\g$ is a generalization of the {\em classical Weil algebra}
\be \Wl =\Sl \otimes \El \label{Wl} \ee
($\ell =$ finite dimensional Lie algebra) which is very well
understood ([GHV]). As a product of symmetric and exterior algebras,
$\Wl$ has a supercommutative associative algebra structure, as well as
two differentials $d$ and $h$ (the classical cohomology and Koszul
differentials respectively) which are inner superderivations. For an
arbitrary $\ell$, we have
\be H(\Wl,h)=H(\Wl,d+h)=K \label{coh1} \ee
($d+h$ is the {\em Weil differential}). If $\ell$ is reductive, it is
also known that
\be H(\Wl,d)=(\Sl)^{\ell}\otimes(\El)^{\ell} \label{coh2} \ee
where superscripts denote subspaces of invariants under the action of
$\ell$ by derivations. It turns out that (\ref{coh2})~is a
{\sl finitely generated} supercommutative associative algebra ([GHV]).
We expect its analogue to be a (finitely generated?) VOSA.
\vsp

It is possible to define $\Wg$ for any tame Lie algebra; B.~Feigin and
E.~Frenkel gave the general definition and computed the semi-infinite
cohomology for the case $\g =Vir$ in [FF]. The next logical choice of
$\g$ is the {\em loop algebra} $\tl$ of a finite dimensional Lie
algebra $\ell$, defined by
\bea && \tl =\ell\otimes K[t,t^{-1}] \nn \\ && {[x\otimes t^{n},
y\otimes t^{m}]}=[x,y]\otimes t^{n+m} \; . \label{loop} \eea
We will call $\Wtl$ a {\em vertex operator Weil algebra} ({\em VOWA})
since it is naturally a VOSA -for what is called the ``standard'', or
``physical'', vacuum- and properly contains the classical algebra
(\ref{Wl}). Even the semi-infinite analogues of $d$ and $h$ (now
called $Q$ and $h$) restrict to (\ref{Wl}) as the classical operators
([A2]). The $Q$-cohomology of the VOWA was first
investigated in [A2].
We will give full details of the semi-infinite theory below.
\vsp

Analogues of $\Sl$ and $\El$, namely the {\em semi-infinite symmetric
and exterior modules} $\Stl$ and $\Etl$, can be defined as induced
modules over the Weyl and Clifford algebras based on $\tl$ with its
restricted dual
\be \tl'=\oplus_nHom(\ell\otimes K\, t^{-n},K) \label{resdual} \ee
respectively. The resulting tensor product module
\be \Wtl =\Stl\otimes\Etl \label{VOWA} \ee
has appeared under the name ``$bc\beta\gamma$-system'' in physics
(although not in connection with semi-infinite cohomology). We will
use the alias ``VOWA'' but keep the notation. Let $\{ u\}$ be a
basis for $\ell$ and $\{ u' \}$ be the dual basis for $\ell'$.
The Clifford algebra on $\ell$ is generated by the symbols
\be b_n^u \; , \; c_n^{u'} \label{bc} \ee
which satisfy the relations
\bea && c_n^{u'}b_m^v+b_m^vc_n^{u'}=<u',v>\delta_{m+n,0}
\nn \\ && b_n^ub_m^v+b_m^vb_n^u=0 \nn \\ && c_n^{u'}c_m^{v'}
+c_m^{v'}c_n^{u'}=0 \, . \label{bcrel} \eea
Similarly the Weyl algebra is generated by
\be \Be_n^u \; , \; \Ga_n^{u'} \label{bega} \ee
with
\bea && \Ga_n^{u'}\Be_m^{v}-\Be_m^{v}\Ga_n^{u'}=<u',v>
\delta_{n+m,0} \nn \\ && \Be_n^u\Be_m^v-\Be_m^v\Be_n^u=0 \nn \\ &&
\Ga_n^{u'}\Ga_m^{v'}-\Ga_m^{v'}\Ga_n^{u'}=0 \; . \label{begarel} \eea
Assigning a superdegree (fermion number) of $-1$ to the $b$'s,
$+1$ to the $c$'s, and
$0$ to the rest of the generators, we will express the relations
(\ref{bcrel}) and (\ref{begarel}) in terms of the supercommutator $[\;
, \; ]$ from this point on. We now designate half of the generators as
{\em creation operators} and the rest as {\em annihilation operators}:
Let
\bea && b_n^u\; , \; \Be_n^u \; \;\mbox{with}\; \; n\leq -1, \; \mbox{and}
\nn \\ && c_n^{u'} \; , \; \Ga_n^{u'} \;\; \mbox{with}\; \; n\leq 0
\label{creation} \eea
be the creation operators. Note that operators of either type
supercommute among themselves. It is possible to shift the division point
(hence choose different {\em vacua}) but then $\Wtl$ will not be a
VOSA. If $\va_{sym}$ and $\va_{ext}$ denote basis elements for the
trivial representations of the subalgebras
generated by the annihilation operators, then $\Stl$ and $\Etl$ are
the induced representations of the Weyl and Clifford algebras, and
$\Wtl$ is spanned by strings of creation operators applied to the
({\em standard}) {\em vacuum}
\be \va=\va_{sym}\otimes\va_{ext} \; . \label{vacuum} \ee
The classical algebra (\ref{Wl}) sits inside (\ref{VOWA}) as the span
of monomials in $c_0^{u'}$ and $\Ga_0^{u'}$.
\vsp

We define an action of the Virasoro algebra on $\Wtl$ via an action on
the (completed) Clifford-Weyl algebra by inner derivations:
This action is independent of the Lie algebra structure of $\ell$ and
reflects the \mbox{$Witt=Der\, K[t,t^{-1}]$} action on the second
factor of (\ref{loop}) ($Vir$ is the unique central extension of
$Witt$ with $L_n$ corresponding to $-t^{n+1}\frac{d}{dt}$). It is easy
to check that
\bea && L_n \cdot b_m^u = -m \, b_{n+m}^u \nn \\ && L_n \cdot \Be_m^u =
-m \, \Be_{n+m}^u \nn \\ && L_n \cdot c_m^{u'} = (-n-m)\,
c_{n+m}^{u'} \nn \\ && L_n \cdot \Ga_m^{u'} = (-n-m)\,
\Ga_{n+m}^{u'} \label{viract} \eea
extends to an action of $Vir$. We have
\be L_n=-\sum_{u,m}m:b_{n+m}^u c_{-m}^{u'}:- \sum_{u,m} m
:\Be_{n+m}^u\Ga_{-m}^{u'}: \label{Ln} \ee
where the infinite sum of products of operators is well-defined on the
module $\Wtl$ and the normal ordering $: \; :$ merely means that an
annihilation operator, if any, should be written on the right (with a
proper sign change). For a more detailed survey of completed
algebras and normal ordering see [A1]. The central element of $Vir$
acts trivially on $\Wtl$, and
\be L_n \cdot \va =0 \Lra n \geq -1. \label{virvac} \ee
Moreover $\Wtl$ is graded by the eigenvalues of $L_0$, as any monomial
is clearly seen to have a nonnegative $L_0$-degree by
(\ref{creation}). The weight zero part is $\Wl$.
\vsp

We need to define a VO $v(z)$ for each basic monomial $\st{v}$.
First let us define ``singles''
\bea && b_{-1}^u \va \, \lra \, b^u(z)=\sum b_n^u z^{-n-1} \nn \\ &&
\Be_{-1}^u \va \, \lra \, \Be^u(z)=\sum \Be_n^u z^{-n-1} \nn \\ &&
c_0^{u'} \va \, \lra \, c^{u'}(z)=\sum c_n^{u'} z^{-n} \nn \\ &&
\Ga_0^{u'} \va \, \lra \, \Ga^{u'}(z)=\sum \Ga_n^{u'} z^{-n} \; .
\label{singles} \eea
Other creation operators are (up to scalars) images of the above
under powers of $L_{-1}$, so by (\ref{dz}), we had better have
\be (b_{-n}^u \va)(z) = \frac{1}{(n-1)!}(\frac{d}{dz})^{n-1} b^u(z),
\ee and so on. For the most general product
\be \st{v}=v^{(1)}\cdots v^{(r)}\va , \; r \geq 1, \label{monomial}
\ee $v^{(i)}=$ a creation operator, we define
\be v(z)= \, : (v^{(1)}\va)(z)\cdots (v^{(r)}\va)(z): \label{field}
\; . \ee
This normal ordering is defined by induction: For $r=2$,
\be :v(z)w(z):=v^{-}(z)w(z)\pm w(z)v^{+}(z), \ee
$v^{\mp}(z)$ indicating half sums over creation and annihilation
operators respectively, and the sign in the middle is $(-)$ only when
both states are odd.
Then (\ref{field}) is defined to be
\be \{ v^{(1)} \}^{-}(z) v^{(2)}(z)\cdots v^{(r)}(z) \pm
v^{(2)}(z)\cdots v^{(r)}(z)\{ v^{(1)} \}^{+}(z). \label{induc} \ee
(In [Li] we have the most general definition of normal ordering as
\be :v(z)w(z):=(v_{-1}\cdot \st{w})(z) \label{bong} \ee
for any $\st{v}$, $\st{w}$ in $V$.) The Virasoro element $\om$
corresponds to the field
\be \omega(z)=-\sum_{u}:b^u(z)\frac{d}{dz}c^{u'}(z):-\sum_u :\Be^u(z)
\frac{d}{dz}\Ga^{u'}(z): \; . \ee
The fact that the
$bc\Be\Ga$-system thus forms a VOSA was known by some experts but
no complete proof was published prior to [A2]. The VOSA structure is
implicit, for example, in [KVDL].
\vsp

{\em 2.2. Semi-infinite Structure and Compatibility}
\vsp

So far we made no use of the specific Lie algebra $\ell$ itself. One
defines an action of $\tl$ on the VOWA via
\be \theta(u_n)=\sum_{v,m}:\Be_{n+m}^{[u,v]}\,\Ga_{-m}^{v'}: +
\sum_{v,m}:b_{n+m}^{[u,v]}\, c_{-m}^{v'}:  \label{totact} \ee
($u_n$ is $u\otimes t^n$).
The well-known formula for the semi-infinite cohomology operator $Q$
([Fe],[A1]) specializes to
\be Q=\sum_{u,v;i,j}:\Be_{i+j}^{[u,v]}\,\Ga_{-i}^{v'}\, c_{-j}^{u'}:
+\sum_{u,v;i<j}:b_{i+j}^{[u,v]}\, c_{-i}^{v'}\, c_{-j}^{u'}:
\label{diff} \; . \ee
The semi-infinite version of the Cartan identity (\ref{Cartan}) is
\be Q\, b_n^u+b_n^u\, Q=\theta(u_n) \; \; \forall u,n. \label{semiCartan}\ee
Once again, we construct all these operators as inner derivations
of the completed Clifford-Weyl algebra ([A2]), and all computations
are reduced to supercommutation relations for the generators, such as
the last equation and
\bea && {[Q,c_n^{u'}]}=-\frac{1}{2}\sum_{v,m}c_{n+m}^{ad'(v)\cdot u'}
\, c_{-m}^{v'} \nn \\ &&
{[Q,\Be_n^u]}=\sum_{v,m}\Be_{n+m}^{[v,u]}\, c_{-m}^{v'} \nn \\ &&
{[Q,\Ga_n^{u'}]}=\sum_{v,m}\Ga_{n+m}^{ad'(v)\cdot u'}\, c_{-m}^{v'}\nn
\; . \label{Qrel} \eea
The superdegree (fermion number) of each operator is easily read
from its formula, by subtracting the number of $b$'s from the number
of $c$'s. The supercommutator is an anticommutator only when both
arguments are odd. We also have
\be Q\cdot \va =0 \ee
and
\be \theta(u_n)\cdot \va =0 \; \Lra \; n\geq 0\, . \ee
Other commutation relations are
\bea && {[Q,\theta(u_n)]}=0 \; \; \forall u,n \nn \\ && {[Q,L_n]}
=0 \; \; \forall n \, . \label{eek} \eea
At this point we also introduce the semi-infinite {\em
Koszul differential}
\be h=\sum_{u,n}\Ga_{-n}^{u'}\, b_n^u \label{Koszul} \ee
([FF]) and a ``homotopy operator''
\be k=\sum_{u,n}n\, \Be_n^u \, c_{-n}^{u'} \ee
which is a modified version of the classical one in [GHV].
\vsp

\begin{prop}\label{hkprop}
The operators $Q$, $h$, $k$ satisfy
\bea && {[Q,h]}={[Q,k]}=0 \nn \\ && {[h,k]}=-L_0 \nn \\ &&
{[h,L_n]}=0 \; \; \forall n \nn \\ && {[k,L_0]}=0
\label{hk} \\ && {[h,\theta(u_n)]}=0 \; \; \forall u,n \nn \\ &&
{[k,\theta(u_0)]}=0 \; \; \forall u \nn \\ &&
Q^2=h^2=k^2=0 \, .  \eea
\end{prop}
\vsp

\begin{prop}
The restrictions of the given operators to the degree-zero (classical)
subspace of $\Wtl$ are
\bea && Q|_{\deg =0}=d_{classical} \nn \\ && h|_{\deg =0}=
h_{classical} \nn \\ && k|_{\deg =0} =0 \nn \\ && \theta(u_0)
|_{\deg =0}=\theta(u) \, . \eea
\end{prop}
\vsp

Proofs are given in [A2]. Then various cohomologies of $\Wtl$ are as
follows:
\vsp

\begin{prop} \label{clas}
\bea && H(\Wtl |_{\deg =0},Q) =H(\Wl,d) \nn \\ && H(\Wtl,h)=
H(\Wl,h)=K \; \; \mbox{by (\ref{hk})} \nn \\ && H(\Wtl,k)=\Wl \eea
\end{prop}
\vsp

The acyclicity of the complex $(\Wtl,Q+h)$ is shown in [FF] by
spectral sequences. We include an elementary proof.
\vsp

\begin{prop}
\be H(\Wtl,Q+h)=K. \ee
\end{prop}

{\em Proof.} Since $[Q+h,k]=-L_0$, the cohomology is restricted to the
classical part. By Proposition~\ref{clas} and (\ref{coh1}) the result
follows. $\Box$
\vsp

We emphasize that these results generalize to all tame Lie algebras.
In the loop algebra case, in order to utilize the powerful VOSA
techniques in $Q$-cohomology computations,
we write the operators above as modes of certain vertex operators
(the method of conversion is just trial and error). There may be more
than one suitable field, in which case we try to choose one
that is in the kernel of $Q$ or the $\ell$ action, etc. Here are
some :
\be J(z)=\sum_{u,v}:\Be^{[u,v]}(z)\Ga^{v'}(z)c^{u'}(z):
+\frac{1}{2}\sum_{u,v}:b^{[u,v]}(z)c^{v'}(z)c^{u'}(z): \ee
is a familiar sight for mathematical physicists; its residue is $Q$.
The actions of elements of $\tl$ are represented by modes of the
operators
\be \theta^u(z)=\sum_v:\Be^{[u,v]}(z)\Ga^{v'}(z):+\sum_{v}
:b^{[u,v]}(z)c^{v'}(z):=\sum_n\theta(u_n)z^{-n-1} \, . \ee
The operators $h$ and $k$ are $h_0$ and $k_1$ for the following VO's:
\be h(z)=\sum_u\Ga^{u'}(z)b^u(z), \; \; \; k(z)=\sum_u\Be^u(z)
\frac{d}{dz}c^{u'}(z) \, . \ee
Recall that by our convention the corresponding states are denoted by
$\st{J}$, $\st{\theta^{u}}$, $\st{h}$, and $\st{k}$.
\vsp

By Proposition~\ref{clas} we know that the classical part of the
cohomology is always there, and indeed there may not be anything else
(it happens for $Vir$ and affine Kac-Moody algebras; see remarks
in [A1]). On the other hand for an
abelian loop algebra $\theta$ and $Q$ are identically zero and we get
rather a lot of cohomology. What can be said about the $Q$-cohomology
for a general loop algebra $\tl$ outside degree zero?
\vsp

\begin{thm} \label{hkthm}
The states
\bea && \st{h}=\sum_u\Ga_0^{u'}b_{-1}^u \va \\ && \st{k}=\sum_u
\Be_{-1}^uc_{-1}^{u'}\va \eea
are both in $(\Wtl)^{\ell}$ and they correspond to nontrivial
$Q$-cohomology classes. As a result, $H(\Wtl [n],Q)$ is nonzero for
every weight $n\geq 0$.
\end{thm}

{\em Proof.} It is straightforward to check
\be \st{h},\st{k} \in (\Wtl)^{\ell} \cap Ker\, Q \ee
(see [A2]). The first state is nonzero in the cohomology
as $Q$ increases the superdegree by 1 and $j=-1$ is already
the smallest superdegree possible for weight one. This implies $L_0$
-hence $Vir$- acts nontrivially on $H(\Wtl,Q)$. But then $\st{k}$
cannot be exact, either, because $k=k_1$ satisfies the nontrivial
relation
\be hk+kh=-L_0 \, . \ee
Finally, we observe both $\st{h}$ and $\st{k}$ are singular vectors.
Therefore we can make use of the relation
\be L_{n}L_{-n}-L_{-n}L_{n}=2n\, L_{0} \ee
to conclude that $L_{-n}\cdot \st{h}$ with weight $n+1$ represents a
nonzero class for each $n\geq 1$. $\Box$
\vsp

\begin{cor}
The Virasoro element
\be \om=-\sum_ub_{-1}^uc_{-1}^{u'}\va -\sum_u\Be_{-1}^u
\Ga_{-1}^{u'}\va \ee
represents a nonzero $Q$-cohomology class.
\end{cor}

{\em Proof.} From (\ref{eek}) we know $[Q,L_n]=0$ for all $n$, which
is equivalent to $Q\cdot\om =0$ by (\ref{rescond}).
The mode $L_0$ of $\omega(z)$ is nonzero in the subquotient VOSA
by the Theorem, hence $\om$ is not $Q$-exact. $\Box$
\vsp

{\bf 3. The Semi-infinite Cohomology for the Semisimple Case}
\vsp

We will analyze the cohomology space $H(\Wtl,Q)$ for a semisimple Lie
algebra $\ell$ using both linear and VOSA methods, and produce
infinitely many explicit nonzero cohomology classes in deg$\neq 0$
for $\ell=sl(2,K)$.
\vsp

Even for an arbitrary tame Lie algebra $\g$, $\Wg$ is a direct sum of
finite dimensional subspaces stable under $Q$ and $\g_0$,
namely those spanned by
monomials with fixed $L_0$-degree and fixed symmetric degree (which is
the counterpart of fermion number in a VOWA, i.e.
the number of $\Ga$'s minus the number of $\Be$'s). For the loop
algebra of semisimple $\ell$ we have $\g_0=\ell$ and complete
reducibility follows. One remarkable corollary is that every
$Q$-cohomology class is representable by an $\ell$ invariant state
just like the classical case (see [GHV]).
\vsp

Recall that $H(\Wtl |_{\deg =0},Q)$ is an associative algebra
generated by finitely many $\ell$ invariants. It is easy to see that
the fields corresponding to the classical generators suffice to
produce all the deg$=0$ cohomology ($c$'s and $\Ga$'s supercommute).
\vsp

{\em 3.1. New Square-zero Operators}
\vsp

The maps $h$ and $k$ make use of the natural correspondence between
the $b$-$\Be$ and $c$-$\Ga$ systems. In case of semisimplicity there
is one more correspondence to be exploited, namely the (Killing)
isomorphism
\be \phi : \ell \ra \ell' \ee
between the $b$-$\Ga$ and $\Be$-$c$ systems.
We add to our list the maps ([A2])
\be r=\sum_{u,n}n\, \Ga_{-n}^{\phi(u)}c_{n}^{u'} \ee
and
\be t=\sum_{u,n}\Be_{-n}^{\phi^{-1}(u')}b_n^u \ee
where
\be r=-r_0, \; \; t=t_1 \ee
for the fields
\be r(z)=\sum_u\Ga^{\phi(u)}(z)\frac{d}{dz}c^{u'}(z)\; \; \mbox{and}
\; \;t(z)=\sum_u\Be^{\phi^{-1}(u')}(z)b^u(z) \, . \ee
The states $\st{h}$, $\st{k}$, $\st{r}$, $\st{t}$ are all
Virasoro-singular reduction elements (they are killed by $L_n$ with
$n>0$). Singular fields are also called ``primary''.
More supercommutators follow.
\vsp

\begin{prop}\label{semicom}
\bea && r^2=t^2=0 \\ && {[r,t]}=-L_0 \\ && {[Q,r]}=[Q,t]=0 \\ &&
{[r,h]}=[r,k]=[t,h]=[t,k]=0  \\ && {[t,\theta(u_n)]}=0 \; \forall
u,n \\ && {[r,\theta(u_0)]}=0 \; \forall u \\ && {[r,L_n]}=0 \;
\forall n \\ && {[t,L_0]}=0 \, . \eea
\end{prop}

{\em Proof.} See [A2]. $\Box$
\vsp

By Propositions \ref{semicom} and \ref{hkprop} we have a decomposition
\be (Ker \, h \cap Ker \, r)\oplus(Ker \, h \cap Ker \,t)\oplus(Ker \,
k \cap Ker \, r)\oplus(Ker\, k \cap Ker \, t) \label{decom} \ee
of each subspace $(\Wtl)|_{\deg =\lambda}$ and its $Q$-cohomology.
This is a useful computational tool, especially since the four
subspaces are isomorphic via maps that commute with $Q$ and $\ell$.
\vsp

We have, in addition to Theorem~\ref{hkthm},
\begin{thm}
The states
\bea && \st{r}=\sum_u \Ga_0^{\phi(u)}\, c_{-1}^{u'}\va \nn \\ &&
\st{t}=\sum_u \Be_{-1}^{\phi^{-1}(u')}\, b_{-1}^u\va \eea
are both invariant under $\ell$ (a semisimple Lie algebra)
and they represent nontrivial $Q$-cohomology classes.
\end{thm}
\vsp

{\em 3.2. The case $sl(2,K)$}
\vsp

Let $sl(2,K)=<x,y,h>$, where
\be {[x,y]}=h, \; \; [h,x]=2x, \; \;\mbox{and}\; \; [h,y]=-2y. \ee
Then basis elements of $\tilde{sl}(2,K)$ will be denoted by $x_n$,
$y_n$, and $h_n$. It is unfortunately more or less
standard to denote both the Koszul differential and the basis element
in the Cartan subalgebra of $sl(2,K)$ by $h$.
\vsp

The coadjoint action of $\tl =\tilde{sl}(2,K)$ on $\tl'$ is given by
the table
\be \begin{array}{cccc} \st{ad'} & \st{x_{r}'} & \st{y_{r}'} &
\st{h_{r}'} \\ \st{x_{m}} & 2h_{r-m}' & 0 & -y_{r-m}' \\
\st{y_{m}} & 0 & -2h_{r-m}' & x_{r-m}' \\ \st{h_{m}} & -2x_{r-m}'
& 2y_{r-m}' & 0. \end{array} \ee
The Killing form ${\cal K} :\ell \times \ell \ra K$ is given by
\be \begin{array}{llll} \st{\Ki} & \st{x} & \st{y} & \st{h} \\
\st{x} & 0 & \frac{1}{2} & 0 \\ \st{y} & \frac{1}{2} & 0 & 0 \\
\st{h} & 0 & 0 & 1. \end{array} \ee
Then $\Ki$ gives rise to the $\ell$-invariant isomorphisms
\be \begin{array}{ll} \phi :\ell \ra \ell', & \phi^{-1}:\ell' \ra \ell
\\ x \mt \frac{1}{2}y' & x' \mt 2y \\ y \mt \frac{1}{2} x' & y' \mt 2x
\\ h \mt h' & h' \mt h. \end{array} \ee

Let us appeal to some classical invariant theory at this stage.
As an $\ell$ module, $\Wtl$ is nothing but the tensor product of
the symmetric and exterior algebras on countably many copies of $\ell$
and $\ell'$, where we distinguish between adjoint and coadjoint
representations, factors to be symmetrized and antisymmetrized, and
factors with different $L_0$-degrees just by looking at the symbols
and their subscripts. The Lie action doesn't change any of these
subdivisions. To obtain all the associative algebra generators for the
$\ell$ invariant subalgebra, all we need to know is the generators of
\be {[S(V)\otimes\wedge(W)]}^{sl(2)} \label{mmm} \ee
where $V$ and $W$ denote direct sums of finitely many copies of, say,
the adjoint representation. By one of those lucky coincidences that
happen at low dimensions the classical $SO(3)$ theory of invariants
tells us all ([Ho],[A2]). The generators of (\ref{mmm}) are quadratic
and cubic, and there is a list of {\sl thirty} types of infinitely
many such generators for the whole invariant subalgebra. We study the
corresponding list of VOSA generators and a miracle follows the lucky
coincidence.
\vsp

\begin{thm} Among the infinitely many purely symmetric $sl(2)$
invariant quadratics and cubics (i.e. those involving only creation
operators in
$\Be$ and $\Ga$'s), only the following five are in $Ker \, Q$:

\bea v^{(1)}\va &=& \{ (\Be_{-1}^h)^2+4\Be_{-1}^x\Be_{-1}^y\} \va \\
v^{(2)}\va &=& \{ \Be_{-1}^h\Be_{-2}^h+2\Be_{-1}^x\Be_{-2}^y+
2\Be_{-1}^y\Be_{-2}^x\} \va \\ v^{(3)}\va &=& \{ (\Ga_0^{h'})^2
+\Ga_0^{x'}
\Ga_0^{y'}\}\va \\ v^{(4)}\va &=& \{ \Ga_0^{h'}\Ga_{-1}^{h'}+
\frac{1}{2}\Ga_0^{x'}
\Ga_{-1}^{y'}+\frac{1}{2}\Ga_0^{y'}\Ga_{-1}^{x'}\}\va \\
v^{(5)}\va &=& \{ \Be_{-1}^h
\Ga_0^{h'}+\Be_{-1}^x\Ga_0^{x'}+\Be_{-1}^y\Ga_0^{y'}\} \va . \eea
None of the $v^{(i)}\va$ is $Q$-exact.
\end{thm}

{\em Proof.} First part of the proof is by direct computation and can
be found in [A2]. The second assertion follows from
\bea h\cdot v^{(1)}\va &=& 2\st{t}\neq 0 \\ L_1 \cdot v^{(2)}\va
&=& 2v^{(1)}\va \neq 0 \\ v^{(3)}\va &=& \mbox{classical}
\\ k\cdot v^{(4)}\va
&=& -\st{r}\neq 0 \\ h\cdot v^{(5)}\va &=& \st{h}\neq 0 \eea
(in the cohomology). $\Box$

This drastic reduction gives us hope for a manageable cohomology. It
is not difficult to show that the following elements are also closed:
\bea && \{ v^{(1)} \}^m \{ v^{(2)} \}^n \va, \; \; m,n \geq 0,\label{c24} \\
&& \{ v^{(3)} \}^m \{ v^{(4)} \}^n \va, \; \; m,n \geq 0,  \\
&& \{ v^{(1)} \}^m v^{(5)} \va, \; \; m \geq 0,  \\
&& \{ v^{(3)} \}^m v^{(5)}  \va, \; \; m \geq 0. \eea

Let $(l,j)$ denote the symmetric and super degrees respectively. We
have the following specific classes for deg$=0$ and deg$=1$: The
classical part consists of classes
\be \{ (\Ga_0^{h'})^2+\Ga_0^{x'}\Ga_0^{y'}\}^n \{ c_0^{h'}c_0^{x'}
c_0^{y'}\}^a \va \label{3ep} \ee
with $n\geq 0$ and $a=0$ or $1$. In particular, there is one
dimensional cohomology for every pair $(l,j)$ with $l\geq 0$, even,
and $j=0$ or $3$. All four operators $h$, $k$, $r$, $t$ act by zero
here as they shift $l$ by $\pm 1$. As for deg$=1$, where it is only
possible to get $j=-1$, $0$, $1$, $2$, $3$, or $4$, we can write the
following infinitely many classes for the first three values of $j$:
\vsp

\begin{thm}
The element
\be \{ v^{(3)}\}^nv^{(4)}\va \ee
is in $Ker \, Q$, as well as $Ker\, h$ and $Ker \, r$, for all $n\geq
0$. Its isomorphic images in the remaining three subspaces in the
decomposition (\ref{decom}) are
\be \{ v^{(3)} \}^n\st{h}, \; \; \{ v^{(3)} \}^n v^{(5)}\va ,\; \;
\mbox{and} \; \; \{ v^{(3)} \}^n \st{r} . \ee
None of the above four is exact.
\end{thm}

{\em Proof.} [A2]. $\Box$
\vsp

The first three types of classes in the Theorem account for all of the
$j=-1$ and $j=0$ cohomology up to $l=10$ (dimensions checked by
computer).
\vsp

{\bf 4. Topological Chiral Algebras}
\vsp

{\em 4.1. Definition and Examples}
\vsp

We take the following definition from [LZ].
A {\em topological chiral algebra} ({\em TCA}) consists of

(i) A VOSA $V$,

(ii) A weight one even field $F(z)$ whose residue (charge) $F_0$
is the ``fermion number operator'',

(iii) A weight one primary (Virasoro-singular) field $J(z)$ with
fermion number one and a square-zero charge $J_0=Q$,

(iv) A weight two primary field $G(z)$ with fermion number $-1$ and
satisfying
\be {[Q,G(z)]}=L(z) \label{a} \ee
where $L(z)$ is the stress-energy field.
\vsp

Examples of TCA's are abundant among semi-infinite cohomological
complexes associated to the Virasoro algebra (see e.g. [LZ]). It
should be pointed out that the above definition is tailored for these
examples: The semi-infinite exterior module for $Vir$ is a simple
$bc$-system generated by two fields $b(z)$ and $c(z)$, and the total
Lie algebra action and the Virasoro action on the VOSA are one and the
same. Eqn. (\ref{a}) is just the Cartan identity in disguise, with
$G(z)=b(z)$. This identity always implies that the tame Lie algebra
acts trivially in the $Q$-cohomology. In case of $Vir$, it also says
the $\om$ element of the theory is exact.
\vsp

{\em 4.2. The VOWA as a TCA}
\vsp

It is obvious from Corollary 1 that VOWA's do not enjoy the above
characteristics. Since $\om$ is not exact, we look for
differentials other than $Q$ to reduce the complex. It turns out
that we have one for every VOWA and an additional one for
semisimple $\ell$.
\vsp

\begin{prop}
\be {[h,k(z)]}=-\omega(z) \ee
and
\be {[r,t(z)]}=-\omega(z). \ee
\end{prop}

{\em Proof.}
\be {[h,k(z)]}=[h_0,k(z)]=(h_0\cdot \st{k})(z) \ee
and
\be {[r,t(z)]}=[r_0,t(z)]=(r_0\cdot \st{t})(z), \ee
but
\be h_0\cdot \st{k}=r_0\cdot \st{t}=-\om \, .\; \;  \Box \ee
\vsp

The fermion numbers of $h$ and $r$ are $-1$ and $1$ respectively. The
field $F(z)$ exists, as in any respectable theory, and is given by
\be F(z)=\sum_u:c^{u'}(z)b^u(z):\, . \ee
Its charge $F_0$ counts the superdegrees.
\vsp

What is the reduction in each case? The $h$-cohomology has already
been identified as $K$. The $r$-cohomology is in degree zero
by virtue of the homotopy relation
\be rt+tr=-L_0 \, , \ee
and for $\ell =sl(2)$ it can be shown to be the one dimensional space
spanned by
\be c_0^{h'}c_0^{x'}c_0^{y'}\va \, . \ee
It may be possible to construct differentials with larger
cohomologies, still necessarily confined to weight zero.
\vsp

{\em 4.3. The Big Picture: Say Cheese!}
\vsp

\be\begin{array}{clclc}
\mbox{QFT} & \la & \mbox{CQFT} & {} & {} \\ \ua & {} & \ua
& {} & {} \\ \mbox{2-D} \; \; \mbox{CFT} & \la & \mbox{TCFT} & \ra
& \mbox{2-D} \; \; \mbox{TFT} \\
\da & {} & \da & {} & \da \\ \mbox{VOSA} & \la & \mbox{TCA} & {}
& \mbox{DT}
\end{array}\ee
\vsp

\begin{center}
{\small\parbox{5in}{ QFT~$=$~Quantum Field Theory,
CQFT~$=$~Cohomological Quantum Field Theory, CFT~$=$~Conformal Field
Theory, TCFT~$=$~Topological Conformal Field Theory,
TFT~$=$~Topological Field Theory, VOSA~$=$~Vertex Operator
Superalgebra, TCA~$=$~Topological Chiral Algebra, DT~$=$~Differential
Topology. Arrows roughly indicate inclusions. Some of the notions are
mathematically rigorous and some are not. ({\em Courtesy}: G.~J.
Zuckerman)}}
\end{center}
\vsp

For more information on the middle column, we recommend [W1], [W2],
[LZ], [DVV], and [Ge].
\vsp

{\bf 5. Conclusion}
\vsp

The VOWA provides a variety of interesting phenomena and is arguably
the simplest VOSA in which we can study them. It has links to both
mathematical physics and to classical Lie algebra theory, including
the theory of invariants. By analogy to classical results we
conjecture that $H(\Wtl,Q)$ for semisimple $\ell$ is finitely
generated as a VOSA, which is supported by the evidence in the special
case $\ell =sl(2)$: The fields $h(z)$, $k(z)$, $r(z)$, $t(z)$, and
$\Pi c^{u'}(z)$ account for all the cohomology classes computed by
direct or indirect methods so far. Producing a canonical set of
generators for any $\ell$ is the ultimate goal in this direction. Also
the $Vir$ module structure of $H(\Wtl,Q)$ and the cohomological
dimensions are yet to be calculated explicitly. What are the primary
fields? The reduction elements? Can we choose generators to be
singular reduction elements (like the five fields above)? Why would we
want to do {\sl that}? Are there any other TCA structures? Eventually,
we would like to attach a (pseudo) physical meaning to the VOWA. The
context may well be topological conformal field theories.
\vsp

{\small{\em Acknowledgements.} I am indebted to my thesis adviser, Gregg
Zuckerman, for his invaluable guidance in every step of this project.
Thanks are also due Bong Lian, Igor Frenkel, Chongying Dong, and Jim
Lepowsky for their help with VOSA's and Roger Howe for his
contribution on invariant theory. I have been partially supported by
Yale University fellowships, NSF grants, an Alfred P. Sloan Foundation
Dissertation Fellowship, and Conference funds.}
\vsp

\frenchspacing

REFERENCES.
\vsp

[A1] F. AKMAN, A characterization of the differential in
semi-infinite cohomology, to appear in {\em J. of Algebra} (preprint
hepth/9302141).
\vv

[A2] F. AKMAN, ``The semi-infinite Weil complex of a graded Lie algebra'',
Yale University thesis, 1993.
\vv

[DVV] R. DIJKGRAAF, H. VERLINDE, AND E. VERLINDE, Notes on topological
string theory and 2d quantum gravity, preprint PUPT-1217,
IASSNS-HEP-90/80.
\vv

[DL] C. DONG AND J. LEPOWSKY, ``Generalized Vertex Algebras and
Relative Vertex Operators'', to appear.
\vv

[Fe] B. FEIGIN, The semi-infinite cohomology of Kac-Moody and
Virasoro Lie algebras, {\em Russ. Math. Surv.} {\bf 39} (1984), 155-156.
\vv

[FF] B. FEIGIN AND E. FRENKEL, Semi-infinite Weil complex and the
Virasoro algebra, {\em Comm. Math. Phys.} {\bf 137} (1991), 617-639.
Erratum: Preprint, February 1992.
\vv

[FGZ] I. B. FRENKEL, H. GARLAND, AND G. J. ZUCKERMAN, Semi-infinite
cohomology and string theory, {\em Proc. Natl. Acad. Sci. USA}
{\bf 83} (1986), 8842-8846.
\vv

[FHL] I. B. FRENKEL, Y.-Z. HUANG, AND J. LEPOWSKY, On axiomatic
approaches to vertex operator algebras and modules, {\em Memoirs AMS}
(1992).
\vv

[FLM] I. B. FRENKEL, J. LEPOWSKY, AND A. MEURMAN, ``Vertex Operator
Algebras and the Monster'', Academic Press, New York, 1988.
\vv

[FZ] I. B. FRENKEL AND Y. ZHU, Vertex operator algebras associated
to representations of affine and Virasoro algebras, {\it Duke Math.
J.} {\bf 66} (1992), 123-168.
\vv

[Ge] E. GETZLER, Batalin-Vilkovisky algebras and two-dimensional
topological field theories, to appear in {\em Comm. Math. Phys.}
(preprint hepth/9212043).
\vv

[GHV] W. GREUB, S. HALPERIN, AND R. VANSTONE, ``Connections,
Curvature, and Cohomology'', Vol.3, Academic Press, New York, 1972-1976.
\vv

[Ho] R. HOWE, Remarks on classical invariant theory, {\em
Trans. Amer. Math. Soc.} {\bf 313} (1989), 539-570.
\vv

[KVDL] V. G. KAC AND J. W. VAN DE LEUR, Super boson-fermion
correspondence, {\em Ann. Inst. Fourier, Grenoble} {\bf 37} (1987), 99-137.
\vv

[Li] B. H. LIAN, On the classification of simple vertex operator
algebras, Univ. of Toronto preprint, February 1992.
\vv

[LZ] B. H. LIAN AND G. J. ZUCKERMAN, New perspectives on the
BRST-algebraic structure of string theory, Nov. 1992 preprint
hepth/9211072.
\vv

[W1] E. WITTEN, Cohomological field theories, IAS preprint.
\vv

[W2] E. WITTEN, On the structure of the topological phase of
two-dimensional gravity, {\em Nucl. Phys. B340} (1990), 281-332.

\end{document}